\documentclass[11pt,twocolumn]{article}

\usepackage{graphicx}
\usepackage{algorithm,algorithmic}
\usepackage{amsmath}
\usepackage{multirow}
\usepackage{float}
\usepackage{url}
\usepackage[none]{hyphenat}
\usepackage{authblk}

\usepackage{txfonts,balance}

\usepackage[hyphenbreaks]{breakurl}

\usepackage{afterpage}

\begin{document}
\newcounter{cntr1}
\newcounter{cntr2}
\emergencystretch 3em

\title{Systematic equation formulation for simulation of power electronic
circuits using explicit methods}

\author{Mahesh~B.~Patil}
\affil{Department of Electrical Engineering, Indian Institute of Technology Bombay}

\maketitle

\begin{abstract}
Use of explicit integration methods for power electronic circuits with
ideal switch models significantly improves simulation speed. The PLECS
package\,\cite{plecs}
has effectively used this idea; however, the implementation details
involved in PLECS are not available in the public domain. Recently, a
basic framework, called the ``ELEX" scheme, for implementing explicit methods
has been described\,\cite{elex1}.
A few modifications of the ELEX scheme for efficient handling of inductors and
switches have been presented in \cite{elex2}.
In this paper, the approach presented in \cite{elex2} is further augmented with 
robust schemes that enable systematic equation formulation 
for circuits involving switches, inductors, and transformers. Several examples
are presented to illustrate the proposed schemes.
\end{abstract}

\section{Introduction}
Explicit integration methods are in general not suitable for circuit simulation since
they become unstable when the circuit time constants are small compared to the simulator
time step. However, for several power electronic circuits, it is possible to avoid small
time constants~-- and thereby circumvent the above limitation~-- by treating the switches
in an idealised manner, viz.,
zero resistance when the switch is closed and
infinite resistance when the switch is open.
This approach leads to different circuits, each corresponding to a
specific set of switch states. Simulation is speeded up substantially since each of the
circuits is linear, and the circuit matrix inverse needs to be computed only once for
any given switch configuration. This basic idea has been used in the PLECS
simulator\,\cite{plecs}, leading to a significant speed advantage over conventional
circuit simulators such as SPICE\,\cite{ngspice} and its variants. Owing to the commercial
nature of PLECS, however, the implementation details have not been shared in the public
domain.

Recently, the ``ELEX" scheme was presented\,\cite{elex1}
to address the basic issues involved in implementing explicit
methods for power electronic circuit simulation. It was pointed out in \cite{elex2} that
the treatment of switches and inductors presented in \cite{elex1} has serious limitations,
and an alternative approach involving Element Stamp (ES) and Circuit Topology Dependent (CTD)
equations was described. In this paper, we point out that,
the treatment of switches and inductors in the ES/CTD approach presented in \cite{elex2}
needs to be augmented. We also discuss transformer equations in the context of the ES/CTD
approach. Furthermore, we present systematic procedures to address the various challenges
associated with assembling circuit equations using the ELEX scheme.

The paper is organised as follows. In Sec.~\ref{sec_sw}, we discuss assembly of equations
related to switches and also the special case of isolation within a circuit due to switches
in the off state.
In Sec.~\ref{sec_ind}, we discuss some issues related to inductor circuits.
We present a systematic approach to check if two inductors are
connected in series. We also give a simple procedure to check if an inductor has a conduction
path.
In Sec.~\ref{sec_xfmr}, we discuss equations related to transformers, and present a general
scheme for systematic formulation of equations for circuits involving both transformers
and inductors.
Finally, in Sec.~\ref{sec_conclusions}, we present the conclusions of this work.

\section{Switch circuits}
\label{sec_sw}
In the ELEX scheme\,\cite{elex2} we write the ES equations for a switch $S$ as
\begin{equation}
\begin{array}{cl}
i_S = 0 &{\textrm{if}}~S~{\textrm{is off}}, \\
V_S = V_{\mathrm{on}} &{\textrm{if}}~S~{\textrm{is on}},
\label{eq_sw_1}
\end{array}
\end{equation}
where $V_{\mathrm{on}}$ is the voltage drop across the switch when it is
conducting. Note that $V_S$ and $i_S$ in Eq.~\ref{eq_sw_1} are auxiliary variables
which represent the switch voltage and switch current, respectively. They need to be
related to the circuit currents and voltages using the CTD equations, which will
vary from circuit to circuit. An example with switch branches was considered
in \cite{elex2} to illustrate this approach. However, there are situations which
require additional considerations in handling the switch equations. In the
following, we illustrate these situations through representative examples.
\subsection{Switch loops}
\label{sec_sw_loops}
In the ELEX scheme presented in \cite{elex2}, an example with parallel switch
branches was considered, and consistent sets of equations were presented for
different switch configurations. In some circuits, switches form a closed loop, but
without any switch branches appearing in parallel, and therefore a more
generalised approach is called for. The following examples illustrate this
point.
\subsubsection{Switch circuits: example 1}
\label{sec_sw_ex_1}
\begin{figure}[!ht]
\centering
\scalebox{0.9}{\includegraphics{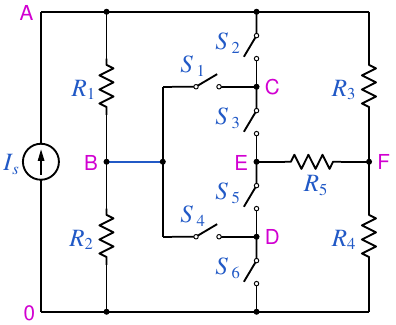}}
\vspace*{-0.2cm}
\caption{Switch circuit example 1.}
\label{fig_sw_1}
\end{figure}

\begin{figure}[!ht]
\centering
\scalebox{0.9}{\includegraphics{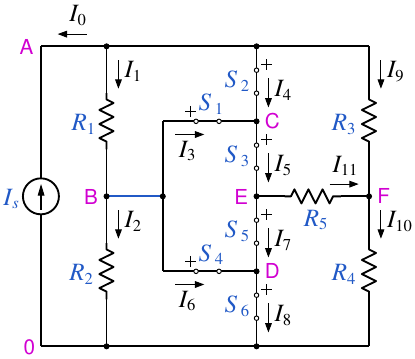}}
\vspace*{-0.2cm}
\caption{Circuit of Fig.~\ref{fig_sw_1} with all switches on.}
\label{fig_sw_1a}
\end{figure}

Consider the circuit shown in Fig.~\ref{fig_sw_1} in which
$S_1$,
$S_3$,
$S_5$,
$S_4$
form a loop of switches. If all switches are on, we get the circuit shown in
Fig.~\ref{fig_sw_1a}. The branch currents assigned by the simulator are shown
in capital letters ($I_0$, $I_1$, etc) in Fig.~\ref{fig_sw_1a} and also in other
figures in the paper. Each switch has nodes $p$ and $n$, with
$V_S \,$=$\, V(p)-V(n)$, where $V(p)$ and $V(n)$ are node voltages with
respect to a reference node (denoted by $0$ in Fig.~\ref{fig_sw_1a}). The switch
current $i_S$ in Eq.~\ref{eq_sw_1} is defined to be positive if the switch
carries a current from the $p$ node to the $n$ node. The $p$ node of each switch
is shown with a $+$ sign. For simplicity, we will consider 
$V_{\mathrm{on}}$ to be $0$ for all switches. Also, we will denote
$V_S$ and $i_S$ for switch $S_1$ by $V_{S1}$ and $i_{S1}$, respectively
(and similarly for other switches).

\begin{figure}[!ht]
\centering
\scalebox{0.9}{\includegraphics{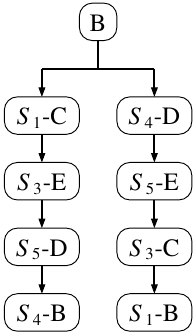}}
\vspace*{-0.2cm}
\caption{Switch tree for identifying switch loops in the circuit of Fig.~\ref{fig_sw_1a},
starting with node B.}
\label{fig_sw_1a_tree}
\end{figure}

The ES equations for the switches are
\begin{equation}
V_{S1} = 0,
\label{eq_sw_ex1_1}
\end{equation}
\begin{equation}
V_{S3} = 0,
\label{eq_sw_ex1_2}
\end{equation}
\begin{equation}
V_{S5} = 0,
\label{eq_sw_ex1_3}
\end{equation}
\begin{equation}
V_{S4} = 0,
\label{eq_sw_ex1_4}
\end{equation}
\begin{equation}
V_{S2} = 0,
\label{eq_sw_ex1_5}
\end{equation}
\begin{equation}
V_{S6} = 0.
\label{eq_sw_ex1_6}
\end{equation}
In addition, we would write the switch voltage definitions (SVD) as
\begin{equation}
V_{S1} = V_B-V_C,
\label{eq_sw_ex1_1a}
\end{equation}
\begin{equation}
V_{S3} = V_C-V_E,
\label{eq_sw_ex1_2a}
\end{equation}
\begin{equation}
V_{S5} = V_E-V_D,
\label{eq_sw_ex1_3a}
\end{equation}
\begin{equation}
V_{S4} = V_B-V_D,
\label{eq_sw_ex1_4a}
\end{equation}
\begin{equation}
V_{S2} = V_A-V_C,
\label{eq_sw_ex1_5a}
\end{equation}
\begin{equation}
V_{S6} = V_D.
\label{eq_sw_ex1_6a}
\end{equation}
Eqs.~\ref{eq_sw_ex1_1}-\ref{eq_sw_ex1_3} and
Eqs.~\ref{eq_sw_ex1_1a}-\ref{eq_sw_ex1_3a} give
$ V_B \,$=$\, V_C \,$=$\, V_D \,$=$\, V_E$, and
Eq.~\ref{eq_sw_ex1_4a} reduces to $V_{S4} \,$=$\, 0$.
However, $V_{S4}$ is already equated to zero by
Eq.~\ref{eq_sw_ex1_4}. Clearly, we need to drop one equation from
Eqs.~\ref{eq_sw_ex1_1a}-\ref{eq_sw_ex1_4a} so that the overall system of
equations is solvable. To this end, we first need to identify the loop
$S_1$-$S_3$-$S_5$-$S_4$, given the netlist for the circuit, i.e., a description
of how the various elements are connected. For this purpose, we will use the
``Switch Loop" (SL) algorithm consisting of the following steps.
\begin{list}{\arabic{cntr1}.}{\usecounter{cntr1}}
 \item
  We call an on switch a ``loop candidate" if it is connected to on switches at
  both ends. In the circuit of Fig.~\ref{fig_sw_1a}, the loop candidates are
  $S_1$, $S_3$, $S_4$, $S_5$. Note that the $p$ node of switch $S_2$ (node A in
  the figure) is not connected to another on switch, and therefore $S_2$ is not
  a loop candidate. Similarly, $S_6$ is not a loop candidate.
 \item
  Start with one of the candidate nodes, and make a ``switch tree". For the
  circuit of Fig.~\ref{fig_sw_1a}, if we start with node B, we get the switch tree
  shown in Fig.~\ref{fig_sw_1a_tree}. From node B, we have two paths:
  (a)\,{\mbox{$S_1$-C}}, i.e., path to node C through $S_1$,
  (b)\,{\mbox{$S_4$-D}}, i.e., path to node D through $S_4$.
  From node C, we have only one path, viz.,
  {\mbox{$S_3$-E}}. Note that, from node C,
  {\mbox{$S_2$-A}} is not a possible path since $S_2$ has not been marked as a
  loop candidate in step\,1. Also, from node C, we do not consider the path
  {\mbox{$S_1$-B}} since $S_1$ has already been traversed.
 \item
  We continue this process until we reach the starting node (B in this case).
  Any branch reaching node B gives us one loop, and the switches involved
  in that loop can be found by tracing back to the root.
\end{list}
\vspace*{-0.4cm}
From Fig.~\ref{fig_sw_1a_tree}, we find that there are two loops starting from
node B:
{\mbox{$S_1$-$S_3$-$S_4$-$S_5$}} (the left branch in the figure) and
{\mbox{$S_4$-$S_5$-$S_3$-$S_1$}} (the right branch). In this case, the two branches
involve the same set of switches, so we conclude that there is only one loop.
We can now drop the SVD for one of the switches, say, Eq.~\ref{eq_sw_ex1_4a}.

The complete set of equations for the circuit of Fig.~\ref{fig_sw_1a} includes
Eqs.~\ref{eq_sw_ex1_1}-\ref{eq_sw_ex1_3a},
\ref{eq_sw_ex1_5a}, \ref{eq_sw_ex1_6a},
and the following additional equations.
\begin{equation}
I_0 + I_1 + I_4 + I_9 = 0,
\label{eq_sw_ex1_KCL_1}
\end{equation}
\begin{equation}
- I_1 + I_2 + I_3 + I_6 = 0,
\label{eq_sw_ex1_KCL_2}
\end{equation}
\begin{equation}
- I_3 - I_4 + I_5 = 0,
\label{eq_sw_ex1_KCL_3}
\end{equation}
\begin{equation}
- I_5 + I_7 + I_{11} = 0,
\label{eq_sw_ex1_KCL_4}
\end{equation}
\begin{equation}
- I_6 - I_7 + I_8 = 0,
\label{eq_sw_ex1_KCL_5}
\end{equation}
\begin{equation}
- I_9 + I_{10} - I_{11} = 0,
\label{eq_sw_ex1_KCL_6}
\end{equation}
\begin{equation}
i_{S1} = I_3,
\label{eq_sw_ex1_swcur_1}
\end{equation}
\begin{equation}
i_{S2} = I_4,
\label{eq_sw_ex1_swcur_2}
\end{equation}
\begin{equation}
i_{S3} = I_5,
\label{eq_sw_ex1_swcur_3}
\end{equation}
\begin{equation}
i_{S4} = I_6,
\label{eq_sw_ex1_swcur_4}
\end{equation}
\begin{equation}
i_{S5} = I_7,
\label{eq_sw_ex1_swcur_5}
\end{equation}
\begin{equation}
i_{S6} = I_8,
\label{eq_sw_ex1_swcur_6}
\end{equation}
\begin{equation}
I_0 = -I_s,
\label{eq_sw_ex1_ES_IS}
\end{equation}
\begin{equation}
V_A-V_B = I_1 R_1,
\label{eq_sw_ex1_ES_R1}
\end{equation}
\begin{equation}
V_B = I_2 R_2,
\label{eq_sw_ex1_ES_R2}
\end{equation}
\begin{equation}
V_A-V_F = I_9 R_3,
\label{eq_sw_ex1_ES_R3}
\end{equation}
\begin{equation}
V_F = I_{10} R_4,
\label{eq_sw_ex1_ES_R4}
\end{equation}
\begin{equation}
V_E-V_F = I_{11} R_5,
\label{eq_sw_ex1_ES_R5}
\end{equation}
\begin{equation}
i_{S4} - i_{S5} - i_{S3} - i_{S1} = 0.
\label{eq_sw_ex1_loop}
\end{equation}
Eqs.~\ref{eq_sw_ex1_KCL_1}-\ref{eq_sw_ex1_KCL_6} are from KCL,
Eqs.~\ref{eq_sw_ex1_swcur_1}-\ref{eq_sw_ex1_swcur_6} assign the switch currents for
on switches to the respective branch currents, and
Eqs.~\ref{eq_sw_ex1_ES_IS}-\ref{eq_sw_ex1_ES_R5} are the ES equations.

\begin{figure}[!ht]
\centering
\scalebox{0.9}{\includegraphics{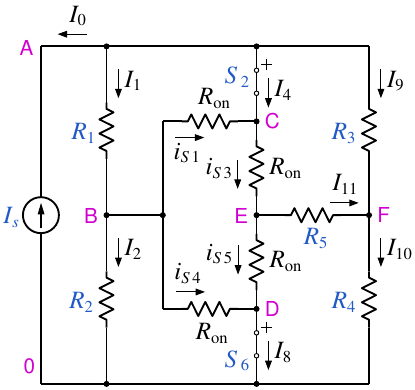}}
\vspace*{-0.2cm}
\caption{Circuit of Fig.~\ref{fig_sw_1a} with loop switches with $R_{\mathrm{on}}$.}
\label{fig_sw_1a_Ron}
\end{figure}

Eq.~\ref{eq_sw_ex1_loop} comes from application of KVL around the loop
{\mbox{B-C-E-D-B}}. If we imagine switch $S_1$ to be represented by a resistance
$R_{\mathrm{on}}$ (see Fig.~\ref{fig_sw_1a_Ron}), the voltage drop across $S_1$ is
$i_{S1} R_{\mathrm{on}}$.
By adding the drops across the four switches in the loop,
we get
\begin{equation}
R_{\mathrm{on}}\,i_{S4} - R_{\mathrm{on}}\,i_{S5} - R_{\mathrm{on}}\,i_{S3} - R_{\mathrm{on}}\,i_{S1} = 0,
\label{eq_sw_ex1_loop_1}
\end{equation}
which reduces to Eq.~\ref{eq_sw_ex1_loop}.

\subsubsection{Switch circuits: example 2}
\label{sec_sw_ex_2}
\begin{figure}[!ht]
\centering
\scalebox{0.9}{\includegraphics{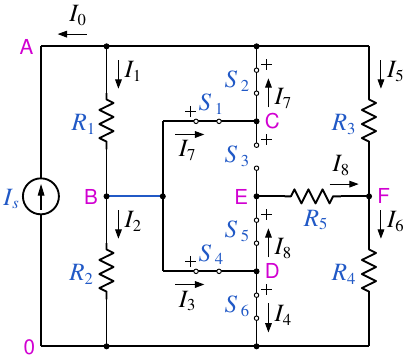}}
\vspace*{-0.2cm}
\caption{Circuit of Fig.~\ref{fig_sw_1} with $S_3$ off and all other switches on.}
\label{fig_sw_2}
\end{figure}

Consider the circuit of Fig.~\ref{fig_sw_1} again with $S_3$ open and all other
switches closed, as shown in Fig.~\ref{fig_sw_2}.
It is clear from the figure that the
{\mbox{$S_1$-$S_3$-$S_5$-$S_4$} loop is broken since $S_3$ is now replaced with
an open circuit. As the first step of the SL algorithm, we identify
$S_1$, $S_4$ as the loop candidates.
Starting with this set, we perform ``passes" as follows.
In the first pass, we observe that one of the on switches $S_1$ is connected to,
viz., $S_2$, is not a loop candidate. We therefore remove
$S_1$ from the candidate set, leaving only $S_4$. In the second pass, we rule out $S_4$ as
well since one of the on switches it connects to, viz., $S_1$, is not a candidate.
That leaves an empty candidate set, and we conclude that no switch loop is possible in this
example.

We now list the equations related to the circuit of Fig.~\ref{fig_sw_2}, skipping the
ES equations for $R_1$-$R_5$ and $I_s$ (which are similar to those seen in
Sec.~\ref{sec_sw_ex_1}).
\begin{equation}
V_{S1} = 0,
\label{eq_sw_ex2_1}
\end{equation}
\begin{equation}
V_{S2} = 0,
\label{eq_sw_ex2_2}
\end{equation}
\begin{equation}
V_{S4} = 0,
\label{eq_sw_ex2_3}
\end{equation}
\begin{equation}
V_{S5} = 0,
\label{eq_sw_ex2_4}
\end{equation}
\begin{equation}
V_{S6} = 0,
\label{eq_sw_ex2_5}
\end{equation}
\begin{equation}
i_{S3} = 0.
\label{eq_sw_ex2_6}
\end{equation}
\begin{equation}
I_0 + I_1 + I_5 - I_7 = 0,
\label{eq_sw_ex2_KCL_1}
\end{equation}
\begin{equation}
- I_1 + I_2 + I_3 + I_7 = 0,
\label{eq_sw_ex2_KCL_2}
\end{equation}
\begin{equation}
- I_3 + I_4 + I_8 = 0,
\label{eq_sw_ex2_KCL_3}
\end{equation}
\begin{equation}
- I_5 + I_6 - I_8 = 0,
\label{eq_sw_ex2_KCL_4}
\end{equation}
\begin{equation}
V_{S1} = V_B - V_C,
\label{eq_sw_ex2_1a}
\end{equation}
\begin{equation}
V_{S2} = V_A - V_C,
\label{eq_sw_ex2_2a}
\end{equation}
\begin{equation}
V_{S4} = V_B - V_D,
\label{eq_sw_ex2_3a}
\end{equation}
\begin{equation}
V_{S5} = V_E - V_D,
\label{eq_sw_ex2_4a}
\end{equation}
\begin{equation}
V_{S6} = V_D,
\label{eq_sw_ex2_5a}
\end{equation}
\begin{equation}
  i_{S1} = I_7,
\label{eq_sw_ex2_swcur_1}
\end{equation}
\begin{equation}
  i_{S2} = -I_7,
\label{eq_sw_ex2_swcur_2}
\end{equation}
\begin{equation}
  i_{S4} = I_3,
\label{eq_sw_ex2_swcur_3}
\end{equation}
\begin{equation}
  i_{S5} = -I_8,
\label{eq_sw_ex2_swcur_4}
\end{equation}
\begin{equation}
  i_{S6} = I_4,
\label{eq_sw_ex2_swcur_5}
\end{equation}
\begin{equation}
  V_{S3} = V_C - V_E.
\label{eq_sw_ex2_off_1}
\end{equation}
Eqs.~\ref{eq_sw_ex2_1}-\ref{eq_sw_ex2_6} are the switch ES equations,
Eqs.~\ref{eq_sw_ex2_KCL_1}-\ref{eq_sw_ex2_KCL_4} are the KCL equations,
Eqs.~\ref{eq_sw_ex2_1a}-\ref{eq_sw_ex2_5a} are the SVDs for on switches,
Eqs.~\ref{eq_sw_ex2_swcur_1}-\ref{eq_sw_ex2_swcur_5} are the branch current
assignments for the on switches, and Eq.~\ref{eq_sw_ex2_off_1} is the
SVD for the off switch $S_3$.

\subsubsection{Switch circuits: example 3}
\label{sec_sw_ex_3}
Consider the circuit shown in Fig.~\ref{fig_sw_3}. When all switches are on,
we have multiple switch loops, viz.,
{\mbox{$S_1$-$S_3$-$S_2$}},
{\mbox{$S_1$-$S_6$-$S_5$-$S_4$}}, and
{\mbox{$S_2$-$S_3$-$S_6$-$S_5$-$S_4$}}
(see Fig.~\ref{fig_sw_3a}).
Note that this case has also been considered in \cite{elex2} (Sec.\,3.1,
example\,1) as an example of three switch branches in parallel. Here, we will
use the more general approach of Sec.~\ref{sec_sw_ex_1} to arrive at the circuit
equations.
\begin{figure}[!ht]
\centering
\scalebox{0.9}{\includegraphics{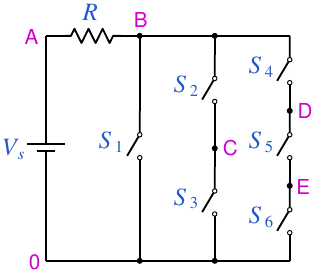}}
\vspace*{-0.2cm}
\caption{Switch circuit example 3.}
\label{fig_sw_3}
\end{figure}
\begin{figure}[!ht]
\centering
\scalebox{0.9}{\includegraphics{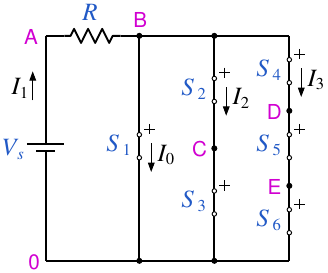}}
\vspace*{-0.2cm}
\caption{Circuit of Fig.~\ref{fig_sw_3} with all switches on.}
\label{fig_sw_3a}
\end{figure}

The SL algorithm starts with identifying potential candidates. Since each switch
in Fig.~\ref{fig_sw_3a} is connected at both ends to on switches, the candidate set
consists of all switches
{\mbox{$S_1$-$S_6$}}.
The switch tree starting with node B is shown in Fig.~\ref{fig_sw_3a_tree}.
\begin{figure}[!ht]
\centering
\scalebox{0.9}{\includegraphics{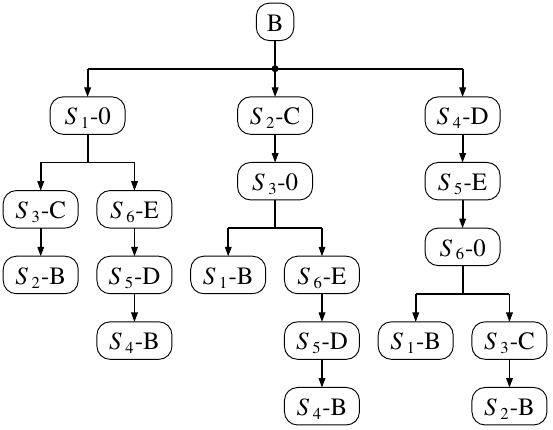}}
\vspace*{-0.2cm}
\caption{Switch tree for identifying switch loops in the circuit of Fig.~\ref{fig_sw_3a},
starting with node B.}
\label{fig_sw_3a_tree}
\end{figure}
We can identify six loops from the tree, but some of them are identical, e.g,
{\mbox{$S_2$-$S_3$-$S_1$}} and
{\mbox{$S_1$-$S_3$-$S_2$}}.
By dropping the repeated loops, we end up with three distinct loops, viz.,
{\mbox{$S_1$-$S_3$-$S_2$}},
{\mbox{$S_1$-$S_6$-$S_5$-$S_4$}}, and
{\mbox{$S_2$-$S_3$-$S_6$-$S_5$-$S_4$}}.

Some caution needs to be exercised in writing the SVD
equations since the loops overlap. For example, loops
{\mbox{$S_1$-$S_3$-$S_2$}} and
{\mbox{$S_1$-$S_6$-$S_5$-$S_4$}}
have $S_1$ in common. As in the example of Sec.~\ref{sec_sw_ex_1}, writing the
SVD equations for all switches will result in an unsolvable system of equations.
We need to drop some of the SVDs such that each of the three loops has at least one SVD
missing. The following set of equations is one such possibility.
\begin{equation}
V_{S1} = V_B,
\label{eq_sw_ex3_1a}
\end{equation}
\begin{equation}
V_{S2} = V_B - V_C,
\label{eq_sw_ex3_2a}
\end{equation}
\begin{equation}
V_{S4} = V_B - V_D,
\label{eq_sw_ex3_3a}
\end{equation}
\begin{equation}
V_{S5} = V_D - V_E.
\label{eq_sw_ex3_4a}
\end{equation}
Next, we write KVL equations for the switch loops as discussed in Sec.~\ref{sec_sw_ex_1}.
However, we need to drop one of the three loops in order to get a solvable system of
equations.
For example, we may drop the loop
{\mbox{$S_2$-$S_3$-$S_6$-$S_5$-$S_4$}} and write the following equations for loops
{\mbox{$S_1$-$S_3$-$S_2$}} and
{\mbox{$S_1$-$S_6$-$S_5$-$S_4$}}, respectively.
\begin{equation}
- i_{S3} - i_{S2} + i_{S1} = 0,
\label{eq_sw_ex3_loop_1}
\end{equation}
\begin{equation}
- i_{S6} - i_{S5} - i_{S4} + i_{S1} = 0.
\label{eq_sw_ex3_loop_2}
\end{equation}
The complete set of equations, apart from
Eqs.~\ref{eq_sw_ex3_1a}-\ref{eq_sw_ex3_loop_2}, is given below.
\begin{equation}
I_0 - I_1 + I_2 + I_3 = 0,
\label{eq_sw_ex3_KCL_1}
\end{equation}
\begin{equation}
i_{S1} = I_0,
\label{eq_sw_ex3_swcur_1}
\end{equation}
\begin{equation}
i_{S2} = I_2,
\label{eq_sw_ex3_swcur_2}
\end{equation}
\begin{equation}
i_{S3} = I_2,
\label{eq_sw_ex3_swcur_3}
\end{equation}
\begin{equation}
i_{S4} = I_3,
\label{eq_sw_ex3_swcur_4}
\end{equation}
\begin{equation}
i_{S5} = I_3,
\label{eq_sw_ex3_swcur_5}
\end{equation}
\begin{equation}
i_{S6} = I_3,
\label{eq_sw_ex3_swcur_6}
\end{equation}
\begin{equation}
V_A = V_s,
\label{eq_sw_ex3_ES_Vs}
\end{equation}
\begin{equation}
V_{S1} = 0,
\label{eq_sw_ex3_ES_S1}
\end{equation}
\begin{equation}
V_{S2} = 0,
\label{eq_sw_ex3_ES_S2}
\end{equation}
\begin{equation}
V_{S3} = 0,
\label{eq_sw_ex3_ES_S3}
\end{equation}
\begin{equation}
V_{S4} = 0,
\label{eq_sw_ex3_ES_S4}
\end{equation}
\begin{equation}
V_{S5} = 0,
\label{eq_sw_ex3_ES_S5}
\end{equation}
\begin{equation}
V_{S6} = 0,
\label{eq_sw_ex3_ES_S6}
\end{equation}
\begin{equation}
V_A-V_B = I_1 R,
\label{eq_sw_ex3_ES_R}
\end{equation}
where Eq.~\ref{eq_sw_ex3_KCL_1} is KCL at node B,
Eqs.~\ref{eq_sw_ex3_swcur_1}-\ref{eq_sw_ex3_swcur_6} are the branch current assignments
for on switches, and
Eqs.~\ref{eq_sw_ex3_ES_Vs}-\ref{eq_sw_ex3_ES_R} are the ES equations.

\subsubsection{Switch circuits: example 4}
\label{sec_sw_ex_4}
We consider the circuit of Fig.~\ref{fig_sw_3} again, with $S_4$ off and all
other switches on (see Fig.~\ref{fig_sw_4}).
\begin{figure}[!ht]
\centering
\scalebox{0.9}{\includegraphics{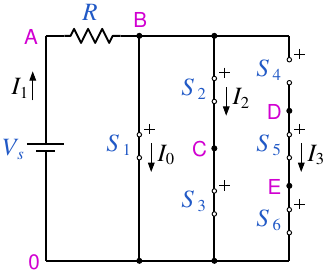}}
\vspace*{-0.2cm}
\caption{Circuit of Fig.~\ref{fig_sw_3} with $S_4$ off and all other switches on.}
\label{fig_sw_4}
\end{figure}
Let us see how the SL algorithm works in this case. We first identify the loop candidates
as $S_1$, $S_2$, $S_3$, $S_6$ since each of them is connected to on switches at both ends.
We then make passes as described in Sec.~\ref{sec_sw_ex_2}. In the first pass, we rule out
$S_6$ which is connected to $S_5$ which is an on switch but not a loop candidate. No other
candidates get ruled out in subsequent passes, and we are left with
$S_1$, $S_2$, $S_3$.

Fig.~\ref{fig_sw_4_tree} shows the switch tree constructed with node C as the root.
From this tree, we identify
{\mbox{$S_1$-$S_3$-$S_2$}} as the switch loop, and choose to omit the SVD equation for
one of them (say, $S_3$).
\begin{figure}[!ht]
\centering
\scalebox{0.9}{\includegraphics{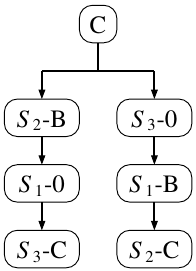}}
\vspace*{-0.2cm}
\caption{Switch tree for identifying switch loops in the circuit of Fig.~\ref{fig_sw_4},
starting with node C.}
\label{fig_sw_4_tree}
\end{figure}
The circuit equations are given by
\begin{equation}
V_{S1} = V_B,
\label{eq_sw_ex4_1a}
\end{equation}
\begin{equation}
V_{S2} = V_B - V_C,
\label{eq_sw_ex4_2a}
\end{equation}
\begin{equation}
V_{S5} = V_D - V_E,
\label{eq_sw_ex4_3a}
\end{equation}
\begin{equation}
V_{S6} = V_E,
\label{eq_sw_ex4_4a}
\end{equation}
\begin{equation}
- i_{S3} - i_{S2} + i_{S1} = 0,
\label{eq_sw_ex4_loop}
\end{equation}
\begin{equation}
I_0 - I_1 + I_2 = 0,
\label{eq_sw_ex4_KCL_1}
\end{equation}
\begin{equation}
I_3 = 0,
\label{eq_sw_ex4_KCL_2}
\end{equation}
\begin{equation}
  i_{S1} = I_0,
\label{eq_sw_ex4_swcur_1}
\end{equation}
\begin{equation}
i_{S2} = I_2,
\label{eq_sw_ex4_swcur_2}
\end{equation}
\begin{equation}
i_{S3} = I_2,
\label{eq_sw_ex4_swcur_3}
\end{equation}
\begin{equation}
i_{S5} = I_3,
\label{eq_sw_ex4_swcur_4}
\end{equation}
\begin{equation}
i_{S6} = I_3,
\label{eq_sw_ex4_swcur_5}
\end{equation}
\begin{equation}
V_{S4} = V_B - V_D,
\label{eq_sw_ex4_off_1}
\end{equation}
\begin{equation}
V_A = V_s,
\label{eq_sw_ex4_ES_Vs}
\end{equation}
\begin{equation}
V_{S1} = 0,
\label{eq_sw_ex4_ES_S1}
\end{equation}
\begin{equation}
V_{S2} = 0,
\label{eq_sw_ex4_ES_S2}
\end{equation}
\begin{equation}
V_{S3} = 0,
\label{eq_sw_ex4_ES_S3}
\end{equation}
\begin{equation}
i_{S4} = 0,
\label{eq_sw_ex4_ES_S4}
\end{equation}
\begin{equation}
V_{S5} = 0,
\label{eq_sw_ex4_ES_S5}
\end{equation}
\begin{equation}
V_{S6} = 0,
\label{eq_sw_ex4_ES_S6}
\end{equation}
\begin{equation}
V_A-V_B = I_1 R,
\label{eq_sw_ex4_ES_R}
\end{equation}
where
Eqs.~\ref{eq_sw_ex4_1a} and \ref{eq_sw_ex4_2a} are the SVD equations for the loop
switches $S_1$ and $S_2$ (note that SVD for $S_3$ has been dropped),
Eqs.~\ref{eq_sw_ex4_3a} and \ref{eq_sw_ex4_4a} are the SVD equations for the other on
switches,
Eq.~\ref{eq_sw_ex4_loop} is the KVL around the
{\mbox{$S_1$-$S_3$-$S_2$}} loop,
Eqs.~\ref{eq_sw_ex4_KCL_1} and \ref{eq_sw_ex4_KCL_2} are KCL equations,
Eqs.~\ref{eq_sw_ex4_swcur_1}-\ref{eq_sw_ex4_swcur_5} are the branch current assignments
for the on switches,
Eq.~\ref{eq_sw_ex4_off_1} is the SVD for the off switch $S_4$, and
Eqs.~\ref{eq_sw_ex4_ES_Vs}-\ref{eq_sw_ex4_ES_R} are the ES equations.

\subsection{Isolated circuit sections}
\label{sec_sw_iso}
In the ELEX scheme presented in \cite{elex2}, equations are
written in terms of node voltages, i.e., voltages with respect to a reference node.
We have used the same approach in the examples considered in this
paper as well. However, in some situations, different sections of the circuit may
get isolated from each other, thus requiring additional considerations, as illustrated in
the following examples.
\subsubsection{Switch circuits: example 5}
\label{sec_sw_ex_5}
Consider the circuit shown in Fig.~\ref{fig_sw_5}. If $S_1$ and $S_4$ are off,
the original circuit gets split into two sections, one on the left and the other on the right,
as shown in Fig.~\ref{fig_sw_5a}.
\begin{figure}[!ht]
\centering
\scalebox{0.9}{\includegraphics{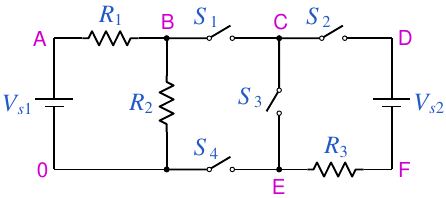}}
\vspace*{-0.2cm}
\caption{Switch circuit example 5.}
\label{fig_sw_5}
\end{figure}
\begin{figure}[!ht]
\centering
\scalebox{0.9}{\includegraphics{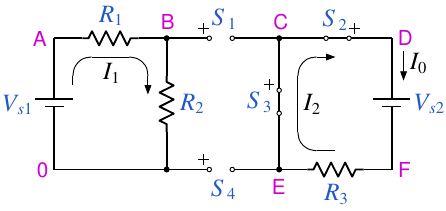}}
\vspace*{-0.2cm}
\caption{Circuit of Fig.~\ref{fig_sw_5} with $S_1$, $S_4$ off and $S_2$, $S_3$ on.}
\label{fig_sw_5a}
\end{figure}
In the left section, the concept
of node voltages is meaningful since the reference node (denoted by 0) is one of the nodes
in this section. The right section, however, is isolated from the reference node, and for
nodes in this section, node voltages would make sense only if we have equations for
the voltage drops across $S_1$ and $S_4$. We can derive the required equations as follows.

We first establish the fact that there is indeed
a section in the circuit (loop {\mbox{C-D-F-E}}) which is isolated from the reference node.
We then identify the off switches responsible for the isolation and obtain
the desired equations. These steps are described below.
\begin{list}{\arabic{cntr1}.}{\usecounter{cntr1}}
 \item
  Identify nodes which are isolated from the reference node (0) by making passes: In the first
  pass, we find that $V_{s1}$ and $R_2$ are connected directly to 0 (see Fig.~\ref{fig_sw_5a}),
  so we tick these elements.
  In the second pass, we look for unticked elements which are connected to any of the ticked elements.
  We find $R_1$ in this category and tick it. In the third pass, we do not find any unticked elements
  connected to a ticked element~-- since $S_1$ and $S_4$ are open~-- and we end the passes. That
  leaves $S_2$, $S_3$, $V_{s2}$, $R_3$ unticked. In this example, the unticked elements happen to
  form a single group since they are connected; however, the simulator in general must take into
  account the possibility of multiple groups isolated from the reference node and from each other.
  The nodes in the group of unticked elements are C, D, F, E. We will refer to these as ``open"
  nodes.

  Note that the simulator has assigned two branch currents, $I_0$ and $I_2$ for the elements
  in the isolated group in Fig.~\ref{fig_sw_5a} although a single branch current could have
  been used. This is related to the branch current assignment section of the simulator and
  will be described elsewhere.
 \item
  Identify off switches for which one of the nodes is an open node: In the example being
  discussed, we look for off switches connected to any of the open nodes (C, D, F, E) and
  find that $S_1$ and $S_4$ satisfy this condition.
 \item
  Write the KCL equation for the off switches identified in the previous step: To arrive at
  a suitable equation, we replace each of $S_1$ and $S_4$ with a large resistance $R_{\mathrm{off}}$,
  as shown in Fig.~\ref{fig_sw_5b}.
  Because the left and right sections of the circuit are isolated, we must have
  \begin{equation}
  i_{S1} + i_{S4} = 0,
  \end{equation}
  leading to
  \begin{equation}
  \displaystyle\frac{V_{S1}}{R_{\mathrm{off}}} +
  \displaystyle\frac{V_{S4}}{R_{\mathrm{off}}} = 0,
  \end{equation}
  which simplifies to
  \begin{equation}
  V_{S1} + V_{S4} = 0.
  \end{equation}
\end{list}
\vspace*{-0.4cm}
\begin{figure}[!ht]
\centering
\scalebox{0.9}{\includegraphics{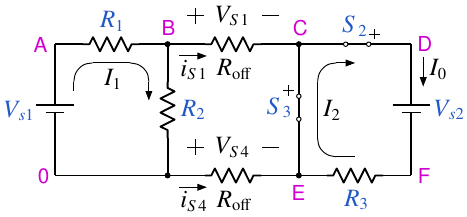}}
\vspace*{-0.2cm}
\caption{Circuit of Fig.~\ref{fig_sw_5a} with $S_1$, $S_4$ replaced with $R_{\mathrm{off}}$.}
\label{fig_sw_5b}
\end{figure}
The rest of the equations are
\begin{equation}
I_0 - I_2 = 0,
\label{eq_sw_ex5_KCL_1}
\end{equation}
\begin{equation}
V_{S2} = V_D - V_C,
\label{eq_sw_ex5_1a}
\end{equation}
\begin{equation}
V_{S3} = V_C - V_E,
\label{eq_sw_ex5_2a}
\end{equation}
\begin{equation}
i_{S2} = - I_2,
\label{eq_sw_ex5_swcur_1}
\end{equation}
\begin{equation}
i_{S3} = - I_2,
\label{eq_sw_ex5_swcur_2}
\end{equation}
\begin{equation}
V_{S1} = V_B - V_C,
\label{eq_sw_ex5_off_1}
\end{equation}
\begin{equation}
V_{S4} =  - V_E,
\label{eq_sw_ex5_off_2}
\end{equation}
\begin{equation}
V_A = V_{s1},
\label{eq_sw_ex5_ES_Vs1}
\end{equation}
\begin{equation}
V_D - V_F = V_{s2},
\label{eq_sw_ex5_ES_Vs2}
\end{equation}
\begin{equation}
V_A-V_B = I_1 R_1,
\label{eq_sw_ex5_ES_R1}
\end{equation}
\begin{equation}
V_B = I_1 R_2,
\label{eq_sw_ex5_ES_R2}
\end{equation}
\begin{equation}
V_F-V_E = I_2 R_3,
\label{eq_sw_ex5_ES_R3}
\end{equation}
\begin{equation}
i_{S1} = 0,
\label{eq_sw_ex5_ES_S1}
\end{equation}
\begin{equation}
V_{S2} = 0,
\label{eq_sw_ex5_ES_S2}
\end{equation}
\begin{equation}
V_{S3} = 0,
\label{eq_sw_ex5_ES_S3}
\end{equation}
\begin{equation}
i_{S4} = 0,
\label{eq_sw_ex5_ES_S4}
\end{equation}
where
Eq.~\ref{eq_sw_ex5_KCL_1} is KCL at node D,
Eqs.~\ref{eq_sw_ex5_1a} and \ref{eq_sw_ex5_2a} are the SVD equations for the on switches,
Eqs.~\ref{eq_sw_ex5_swcur_1} and \ref{eq_sw_ex5_swcur_2} are the branch current assignments
for the on switches,
Eqs.~\ref{eq_sw_ex5_off_1} and \ref{eq_sw_ex5_off_2} are the SVD equations for the off switches,
and
Eqs.~\ref{eq_sw_ex5_ES_Vs1}-\ref{eq_sw_ex5_ES_S4} are the ES equations.

\subsubsection{Switch circuits: example 6}
\label{sec_sw_ex_6}
Consider the circuit shown in Fig.~\ref{fig_sw_6}. If $S_1$ and $S_2$ are open,
$R_1$ gets isolated from the reference node
(see Fig.~\ref{fig_sw_6a}).
\begin{figure}[!ht]
\centering
\scalebox{0.9}{\includegraphics{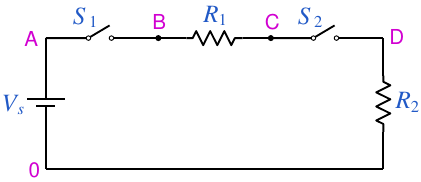}}
\vspace*{-0.2cm}
\caption{Switch circuit example 6.}
\label{fig_sw_6}
\end{figure}
\begin{figure}[!ht]
\centering
\scalebox{0.9}{\includegraphics{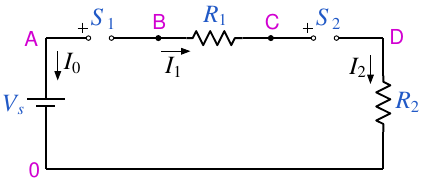}}
\vspace*{-0.2cm}
\caption{Circuit of Fig.~\ref{fig_sw_6} with $S_1$, $S_2$ off.}
\label{fig_sw_6a}
\end{figure}
We identify nodes B and C as open nodes, and find that $S_1$ as well as
$S_2$ is directly connected to an open node. Following the approach described in
the previous section, we can write the KCL equation involving $S_1$ and $S_2$ as
\begin{equation}
-V_{S1} + V_{S2} = 0.
\end{equation}
The rest of the equations can be written as
\begin{equation}
I_0 = 0,
\label{eq_sw_ex6_KCL_1}
\end{equation}
\begin{equation}
I_1 = 0,
\label{eq_sw_ex6_KCL_2}
\end{equation}
\begin{equation}
I_2 = 0,
\label{eq_sw_ex6_KCL_3}
\end{equation}
\begin{equation}
V_{S1} = V_A - V_B,
\label{eq_sw_ex6_off_1}
\end{equation}
\begin{equation}
V_{S2} = V_C - V_D,
\label{eq_sw_ex6_off_2}
\end{equation}
\begin{equation}
V_A = V_s,
\label{eq_sw_ex6_ES_Vs}
\end{equation}
\begin{equation}
V_B-V_C = I_1 R_1,
\label{eq_sw_ex6_ES_R1}
\end{equation}
\begin{equation}
V_D = I_2 R_2,
\label{eq_sw_ex6_ES_R2}
\end{equation}
\begin{equation}
i_{S1} = 0,
\label{eq_sw_ex6_ES_S1}
\end{equation}
\begin{equation}
i_{S2} = 0,
\label{eq_sw_ex6_ES_S2}
\end{equation}
where
Eqs.~\ref{eq_sw_ex6_KCL_1}-\ref{eq_sw_ex6_KCL_3} are the KCL equations,
Eqs.~\ref{eq_sw_ex6_off_1} and \ref{eq_sw_ex6_off_2} are the SVD equations for the off switches,
and
Eqs.~\ref{eq_sw_ex6_ES_Vs}-\ref{eq_sw_ex6_ES_S2} are the ES equations.

\section{Inductor circuits}
\label{sec_ind}
In the ELEX scheme\,\cite{elex2}, inductor equations, like switch equations, are
divided into ES and CTD equations. The ES equations are given by
\begin{equation}
i_L = i_L^{(n+1)},
\end{equation}
\begin{equation}
V_p-V_n - L\,i_{Ld} = 0,
\label{eq_ind_ild}
\end{equation}
where $i_L$ and $i_{Ld}$ are auxiliary variables associated with the inductor,
representing the inductor current and its derivative with respect to time,
respectively, and $V_p$, $V_n$ are the node voltages
$V(p)$, $V(n)$ of the inductor with respect to the reference node.
$i_L^{(n+1)}$
is a {\it{known}} value computed prior to solving the circuit equations.
We will assume for simplicity that the
Forward Euler (FE) scheme is used in handling time derivatives.
As explained in \cite{elex1}, the FE scheme is not a suitable
choice for many circuits of practical interest, and higher-order variable time step methods
such as the Runge-Kutta-Fehlberg (RKF) method are required. However, the nature of the equations
to be solved remains the same in the FE method or RKF method. With the FE scheme, we have
\begin{equation}
i_L^{(n+1)} =
i_L^{(n)} + \displaystyle\frac{\Delta t}{L}\,\left(V_p^{(n)}-V_n^{(n)}\right),
\label{eq_ind_FE}
\end{equation}
where the superscripts $(n)$ and $(n+1)$ correspond to the time points
$t_n$ and $t_{n+1}$, respectively.

The CTD equation for an inductor $L_i$ depends on two crucial factors:
(a)~whether $L_i$ is in series with any other inductor(s),
(b)~whether there is a conduction path available for $L_i$. This information
can be obtained by constructing the graph of the circuit under consideration
and analysing it for series connections and conduction paths. In this paper, we
present much simpler~-- and more robust~-- procedures to achieve the same goal.
\subsection{Inductors in series}
\label{sec_ind_series}
Consider a circuit consisting of inductors $L_i$, $L_j$, along with other elements,
$e_1$, $e_2$, $\cdots$, $e_N$.
For now, assume that each of
$e_1$, $e_2$, $\cdots$, $e_N$ has only two nodes. The following algorithm can be
used to check if
$L_i$ and $L_j$ are in series.
\begin{algorithm}
 \caption{Shorting}
 \begin{algorithmic}[1]
  \FOR {$k$ = 1 to $N$}
   \STATE Short $e_k$.
   \IF{$L_i$ or $L_j$ got shorted}
     \STATE {{\tt{flag\_series = false}}}
     \STATE {exit}
   \ENDIF
  \ENDFOR
  \IF{$L_i$ and $L_j$ are in parallel}
    \STATE {{\tt{flag\_series = true}}}
  \ELSE
    \STATE {{\tt{flag\_series = false}}}
  \ENDIF
 \end{algorithmic}
\end{algorithm}
We keep shorting the elements in the circuit one by one (except for $L_i$ and $L_j$).
If at any stage $L_i$ or $L_j$ (or both) get shorted, we conclude that they are not
in series. After all elements have been shorted, if $L_i$ and $L_j$ have got
connected in parallel, we conclude that they are in series. The following examples
illustrate the working of the Shorting algorithm.
\subsubsection{Inductor circuits: example 1}
\label{sec_ind_ex_1}
Consider the circuit shown in Fig.~\ref{fig_ind_1}. After applying the Shorting algorithm for
{\mbox{($L_1$,\,$L_4$)}}, we get the circuit of
Fig.~\ref{fig_ind_1_L1_L4}, and we find that $L_1$ and $L_4$ have got connected in parallel.
We conclude therefore that $L_1$ and $L_4$ are in series (in the original circuit).

Application of the Shorting algorithm for
{\mbox{($L_1$,\,$L_2$)}}
to the circuit of Fig.~\ref{fig_ind_1} is shown in
Fig.~\ref{fig_ind_1_L1_L2}. In this case, we find that $L_1$ and $L_2$ have both got shorted,
leading to the conclusion that they are not in series. Note that the Shorting algorithm would
in fact stop as soon as any one of
$L_1$, $L_2$ gets shorted, and it would not be required to continue up to the situation shown
in Fig.~\ref{fig_ind_1_L1_L2}.
\begin{figure}[!ht]
\centering
\scalebox{0.9}{\includegraphics{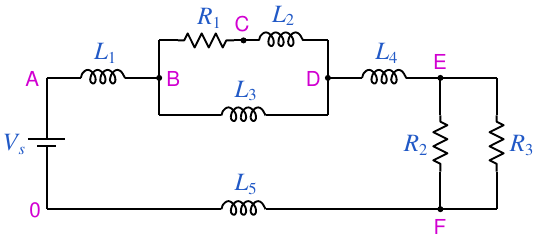}}
\vspace*{-0.2cm}
\caption{Inductor circuit example 1.}
\label{fig_ind_1}
\end{figure}
\begin{figure}[!ht]
\centering
\scalebox{0.9}{\includegraphics{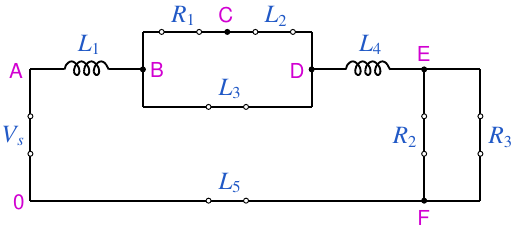}}
\vspace*{-0.2cm}
\caption{Application of the Shorting algorithm to the circuit of Fig.~\ref{fig_ind_1}
for inductors $L_1$, $L_4$.}
\label{fig_ind_1_L1_L4}
\end{figure}
\begin{figure}[!ht]
\centering
\scalebox{0.9}{\includegraphics{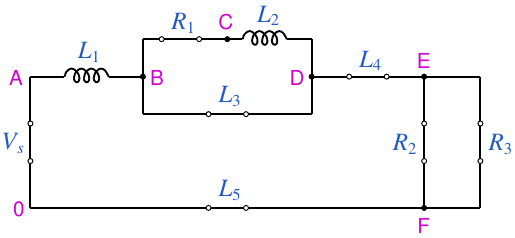}}
\vspace*{-0.2cm}
\caption{Application of the Shorting algorithm to the circuit of Fig.~\ref{fig_ind_1}
for inductors $L_1$, $L_2$.}
\label{fig_ind_1_L1_L2}
\end{figure}

\subsubsection{Inductor circuits: example 2}
\label{sec_ind_ex_2}
If the circuit includes switches, the circuit graph depends on the switch states, and
whether or not $L_i$ and $L_j$ are in series would also depend on the switch states.
Consider the circuit shown in Fig.~\ref{fig_ind_2}. If all switches are on, we get the
circuit shown in Fig.~\ref{fig_ind_2a}. Application of the Shorting algorithm to this
circuit for
{\mbox{($L_1$,\,$L_2$)}}
gives the circuit shown in Fig.~\ref{fig_ind_2a_L1_L2}, and we find $L_1$ and $L_2$ to
be shorted, thus implying that they are not connected in series.

If $S_1$ and $S_3$ are on and $S_2$ off, we get the circuit shown in Fig.~\ref{fig_ind_2b},
and application of the Shorting algorithm for
{\mbox{($L_1$,\,$L_2$)}}
results in the circuit shown in
Fig.~\ref{fig_ind_2b_L1_L2}. We see that $L_1$ and $L_2$ have got connected in parallel
and conclude that they are in series in the original circuit.
\begin{figure}[!ht]
\centering
\scalebox{0.9}{\includegraphics{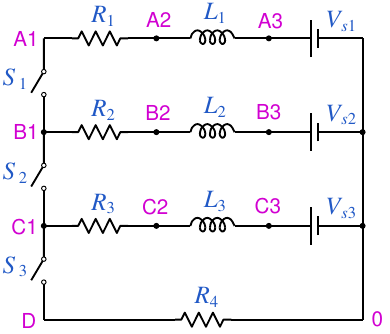}}
\vspace*{-0.2cm}
\caption{Inductor circuit example 2.}
\label{fig_ind_2}
\end{figure}
\begin{figure}[!ht]
\centering
\scalebox{0.9}{\includegraphics{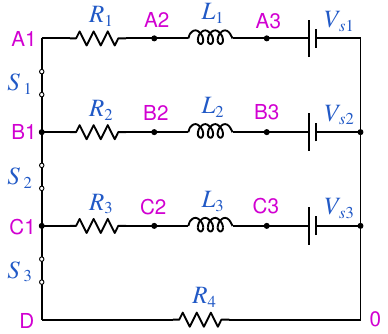}}
\vspace*{-0.2cm}
\caption{Circuit of Fig.~\ref{fig_ind_2} with all switches on.}
\label{fig_ind_2a}
\end{figure}
\begin{figure}[!ht]
\centering
\scalebox{0.9}{\includegraphics{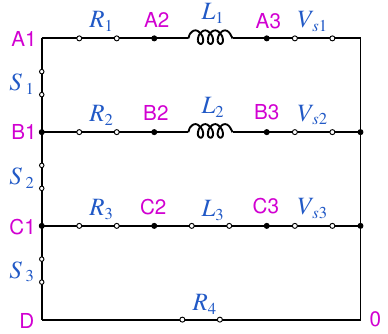}}
\vspace*{-0.2cm}
\caption{Application of the Shorting algorithm to the circuit of Fig.~\ref{fig_ind_2a}
for inductors $L_1$, $L_2$.}
\label{fig_ind_2a_L1_L2}
\end{figure}
\begin{figure}[!ht]
\centering
\scalebox{0.9}{\includegraphics{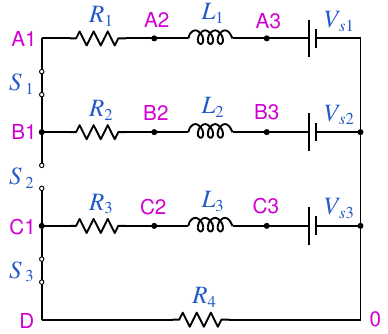}}
\vspace*{-0.2cm}
\caption{Circuit of Fig.~\ref{fig_ind_2} with $S_1$, $S_3$ on, $S_2$ off.}
\label{fig_ind_2b}
\end{figure}
\begin{figure}[!ht]
\centering
\scalebox{0.9}{\includegraphics{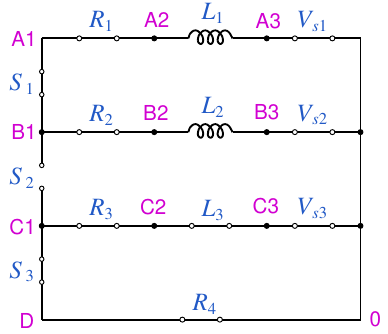}}
\vspace*{-0.2cm}
\caption{Application of the Shorting algorithm to the circuit of Fig.~\ref{fig_ind_2b}
for inductors $L_1$, $L_2$.}
\label{fig_ind_2b_L1_L2}
\end{figure}

\subsection{Conduction path for inductors}
\label{sec_ind_path}
As pointed out in \cite{elex2}, the CTD equation for an inductor depends on whether
there is a conduction path available for that inductor.
Consider a circuit consisting of an inductor $L$ along with other elements,
$e_1$, $e_2$, $\cdots$, $e_N$.
The Path algorithm given below can be used to check if there is a conduction path
for $L$. We keep shorting elements in the circuit~-- except for $L$~-- one by one.
If at any stage $L$ gets shorted, we conclude that $L$ has a conduction path
available, and exit the algorithm. The following examples illustrate the
functioning of the Path algorithm.
\begin{algorithm}
 \caption{Path}
 \begin{algorithmic}[1]
  \STATE {{\tt{flag\_path = false}}}
  \FOR {$k$ = 1 to $N$}
   \STATE Short $e_k$.
   \IF{$L$ got shorted}
     \STATE {{\tt{flag\_path = true}}}
     \STATE {exit}
   \ENDIF
  \ENDFOR
 \end{algorithmic}
\end{algorithm}

\subsubsection{Inductor circuits: example 3}
\label{sec_ind_ex_3}
Consider the circuit of Fig.~\ref{fig_ind_2} with
$S_1$, $S_3$ on and $S_2$ off, which gives the circuit shown in Fig.~\ref{fig_ind_2b}.
This case was considered in Sec.~\ref{sec_ind_ex_2} in the context of checking series
connection of inductors. Application of the Path algorithm for inductor $L_1$ gives
the circuit of Fig.~\ref{fig_ind_2b_L1}. We observe that $L_1$ has got shorted and
conclude that it has a conduction path in the original circuit. Note that the Path
algorithm would terminate as soon as $L_1$ gets shorted, and we do not need to
proceed all the way to the situation shown in Fig.~\ref{fig_ind_2b_L1}.
\begin{figure}[!ht]
\centering
\scalebox{0.9}{\includegraphics{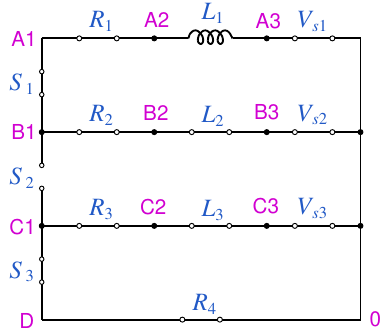}}
\vspace*{-0.2cm}
\caption{Application of the Path algorithm to the circuit of Fig.~\ref{fig_ind_2b}
for inductor $L_1$.}
\label{fig_ind_2b_L1}
\end{figure}
\subsubsection{Inductor circuits: example 4}
\label{sec_ind_ex_4}
Consider the circuit of Fig.~\ref{fig_ind_2} again but with
$S_2$, $S_3$ on and $S_1$ off, leading to the circuit shown in Fig.~\ref{fig_ind_2c}.
Application of the Path algorithm for inductor $L_1$ to this circuit is shown in
Fig.~\ref{fig_ind_2c_L1}. Since $L_1$ has not got shorted, we conclude that it has
no conduction path available in the original circuit.
\begin{figure}[!ht]
\centering
\scalebox{0.9}{\includegraphics{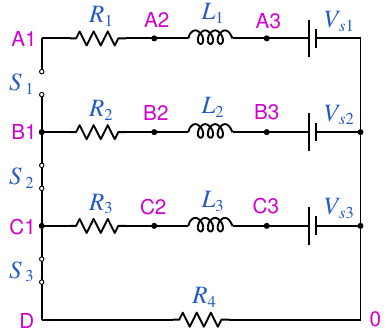}}
\vspace*{-0.2cm}
\caption{Circuit of Fig.~\ref{fig_ind_2} with $S_2$, $S_3$ on, $S_1$ off.}
\label{fig_ind_2c}
\end{figure}
\begin{figure}[!ht]
\centering
\scalebox{0.9}{\includegraphics{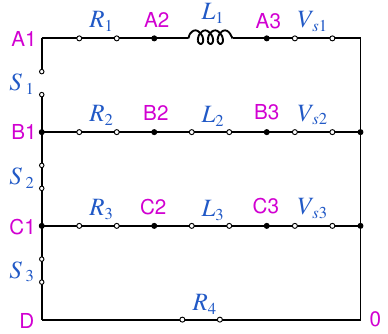}}
\vspace*{-0.2cm}
\caption{Application of the Path algorithm to the circuit of Fig.~\ref{fig_ind_2c}
for inductor $L_1$.}
\label{fig_ind_2c_L1}
\end{figure}

\section{Transformer circuits}
\label{sec_xfmr}
Handling of transformers for circuit simulation using explicit integration methods
presents some challenges, and the situation gets more complicated for circuits
involving transformers, inductors, and switches. In this section, we discuss how
the ELEX scheme of \cite{elex2} can be modified to incorporate transformers. As with
switches and inductors, we divide the equations arising due to transformers into two
categories: ES and CTD.
\subsection{Transformers: ES equations}
\label{sec_xfmr_ES}
Transformer models used in circuit simulation involve an ideal transformer model along
with resistors and inductors to represent non-ideal effects.
\begin{figure}[!ht]
\centering
\scalebox{0.9}{\includegraphics{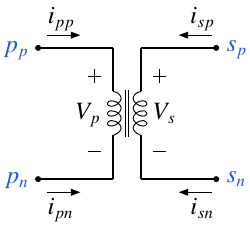}}
\vspace*{-0.2cm}
\caption{Ideal 1:1 transformer.}
\label{fig_xfmr_1_1}
\end{figure}
\begin{figure}[!ht]
\centering
\scalebox{0.9}{\includegraphics{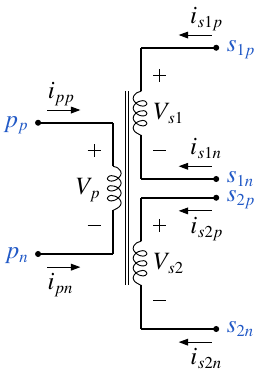}}
\vspace*{-0.2cm}
\caption{Ideal 1:2 transformer.}
\label{fig_xfmr_1_2}
\end{figure}
The ES equations for an ideal 1:1 transformer shown in Fig.~\ref{fig_xfmr_1_1} are
\begin{equation}
V_p = V(p_p)-V(p_n),
\label{eq_xfmr_1_1_Vp}
\end{equation}
\begin{equation}
V_s = V(s_p)-V(s_n),
\label{eq_xfmr_1_1_Vs}
\end{equation}
\begin{equation}
N_s V_p - N_p V_s = 0,
\label{eq_xfmr_1_1_voltage}
\end{equation}
\begin{equation}
N_p i_{pp} + N_s i_{sp} = 0,
\label{eq_xfmr_1_1_current}
\end{equation}
\begin{equation}
i_{pp} + i_{pn} = 0,
\label{eq_xfmr_1_1_kcl_p}
\end{equation}
\begin{equation}
i_{sp} + i_{sn} = 0,
\label{eq_xfmr_1_1_kcl_s}
\end{equation}
where $N_p$ and $N_s$ are the number of turns for the $p$ and $s$ coils, respectively.
The above set of equations can be extended in a straightforward manner to the case of
multiple primary or secondary windings. For example, for a 1:2 transformer (see
Fig.~\ref{fig_xfmr_1_2}), we can write
\begin{equation}
V_p = V(p_p)-V(p_n),
\label{eq_xfmr_1_2_Vp}
\end{equation}
\begin{equation}
V_{s1} = V(s_{1p})-V(s_{1n}),
\label{eq_xfmr_1_2_Vs1}
\end{equation}
\begin{equation}
V_{s2} = V(s_{2p})-V(s_{2n}),
\label{eq_xfmr_1_2_Vs2}
\end{equation}
\begin{equation}
N_{s1} V_p - N_p V_{s1} = 0,
\label{eq_xfmr_1_2_voltage_1}
\end{equation}
\begin{equation}
N_{s2} V_p - N_p V_{s2} = 0,
\label{eq_xfmr_1_2_voltage_2}
\end{equation}
\begin{equation}
N_p i_{pp} + N_{s1} i_{s1p} + N_{s2} i_{s2p} = 0,
\label{eq_xfmr_1_2_current}
\end{equation}
\begin{equation}
i_{pp} + i_{pn} = 0,
\label{eq_xfmr_1_2_kcl_p}
\end{equation}
\begin{equation}
i_{s1p} + i_{s1n} = 0,
\label{eq_xfmr_1_2_kcl_s1}
\end{equation}
\begin{equation}
i_{s2p} + i_{s2n} = 0,
\label{eq_xfmr_1_2_kcl_s2}
\end{equation}
where $N_p$, $N_{s1}$, $N_{s2}$ are the number of turns for the $p$, $s1$, $s2$ coils, respectively.

\subsection{Transformers: CTD equations}
\label{sec_xfmr_CTD}
Transformers introduce significant complications in assembling equations with the ELEX
scheme. Although the primary and secondary sides seem to be independent sections, they
are related through the ideal transformer current equation
(Eq.~\ref{eq_xfmr_1_1_current} for a 1:1 transformer,
Eq.~\ref{eq_xfmr_1_2_current} for a 1:2 transformer)
thus requiring additional considerations while writing the CTD equations. It is perhaps
best to illustrate the challenges posed by transformer circuits with the help of examples.
We start with a simple transformer circuit and then go on to more complicated situations.
\subsubsection{Transformer circuits: example 1}
\label{sec_xfmr_ex_1}
Consider the circuit shown in Fig.~\ref{fig_xfmr_1}.
\begin{figure}[!ht]
\centering
\scalebox{0.9}{\includegraphics{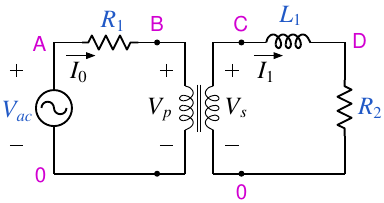}}
\vspace*{-0.2cm}
\caption{Transformer circuit example 1. For $L_1$, nodes $p$ and $n$ are C and D, respectively.}
\label{fig_xfmr_1}
\end{figure}
The variables involved in this circuit
are the branch currents ($I_0$, $I_1$), the node voltages
($V_A$, $V_B$, $V_C$, $V_D$), the transformer variables
($V_p$, $V_s$, $i_{pp}$, $i_{pn}$, $i_{sp}$, $i_{sn}$), and the inductor variables
($i_L$, $i_{Ld}$).
The total number of variables is 14, so we need a set of 14 equations. Note that, since we use
node voltages in the ELEX scheme for formulating the equations, the secondary side is required
to have a connection to the reference node. This is achieved by assigning the same node ($0$) for
the $p_n$ and $s_n$ terminals of the transformer. We can now write the circuit equations as follows.
\begin{equation}
V_p = V_B,
\label{eq_xfmr_ex1_ES_X_1}
\end{equation}
\begin{equation}
V_s = V_C,
\label{eq_xfmr_ex1_ES_X_2}
\end{equation}
\begin{equation}
N_s V_p - N_p V_s = 0,
\label{eq_xfmr_ex1_ES_X_3}
\end{equation}
\begin{equation}
N_p i_{pp} + N_s i_{sp} = 0,
\label{eq_xfmr_ex1_ES_X_4}
\end{equation}
\begin{equation}
i_{pp} + i_{pn} = 0,
\label{eq_xfmr_ex1_ES_X_5}
\end{equation}
\begin{equation}
i_{sp} + i_{sn} = 0,
\label{eq_xfmr_ex1_ES_X_6}
\end{equation}
\begin{equation}
V_A = V_{ac},
\label{eq_xfmr_ex1_ES_Vac}
\end{equation}
\begin{equation}
V_A-V_B = I_0 R_1,
\label{eq_xfmr_ex1_ES_R1}
\end{equation}
\begin{equation}
V_D = I_1 R_2,
\label{eq_xfmr_ex1_ES_R2}
\end{equation}
\begin{equation}
i_{L1} = i_{L1}^{(n+1)},
\label{eq_xfmr_ex1_ES_L1_1}
\end{equation}
\begin{equation}
V_C-V_D = L_1\,i_{L1d},
\label{eq_xfmr_ex1_ES_L1_2}
\end{equation}
\begin{equation}
-I_0 + i_{pp} = 0,
\label{eq_xfmr_ex1_KCL_1}
\end{equation}
\begin{equation}
I_1 + i_{sp} = 0,
\label{eq_xfmr_ex1_KCL_2}
\end{equation}
\begin{equation}
I_1 = i_{L1},
\label{eq_xfmr_ex1_br_L}
\end{equation}
where
Eqs.~\ref{eq_xfmr_ex1_ES_X_1}-\ref{eq_xfmr_ex1_ES_L1_2} are the ES equations,
Eqs.~\ref{eq_xfmr_ex1_KCL_1} and \ref{eq_xfmr_ex1_KCL_2} are the KCL equations at nodes
B and C, respectively,
and Eq.~\ref{eq_xfmr_ex1_br_L} is the branch current assignment for the inductor.

\subsubsection{Transformer circuits: example 2}
\label{sec_xfmr_ex_2}
We now consider a circuit with two inductors, $L_1$ on the primary and $L_2$ on the
secondary side of the transformer, as shown in  Fig.~\ref{fig_xfmr_2}.
\begin{figure}[!ht]
\centering
\scalebox{0.9}{\includegraphics{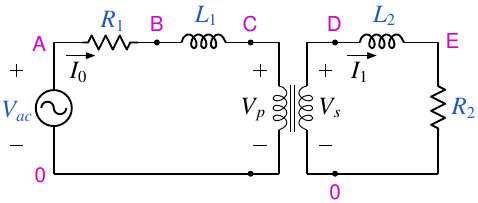}}
\vspace*{-0.2cm}
\caption{Transformer circuit example 2. The $p$ nodes of
$L_1$, $L_2$ are B, D, respectively.}
\label{fig_xfmr_2}
\end{figure}
Although $L_1$ and $L_2$ are separated by the transformer, their currents are linked
since $i_{pp}$ and $i_{sp}$ of the transformer are related by Eq.~\ref{eq_xfmr_1_1_current}.
This situation is similar to Example\,1, Sec.\,2.1, of \cite{elex2}, with two inductors in
series carrying the same current. With the FE scheme, we would write (among other equations)
\begin{equation}
i_{L1} = i_{L1}^{(n+1)},
\label{eq_xfmr_ex2_1}
\end{equation}
\begin{equation}
i_{L2} = i_{L2}^{(n+1)}.
\label{eq_xfmr_ex2_2}
\end{equation}
In addition, the transformer current equation implies that
\begin{equation}
N_p i_{L1} - N_s i_{L2} = 0.
\label{eq_xfmr_ex2_3}
\end{equation}
Clearly,
Eqs.~\ref{eq_xfmr_ex2_1}-\ref{eq_xfmr_ex2_3} cannot be solved since there are two variables and
three equations, and as in the case of inductors in series\,\cite{elex2}, we need to replace
Eq.~\ref{eq_xfmr_ex2_2} with another suitable equation, viz.,
\begin{equation}
N_p i_{L1d} - N_s i_{L2d} = 0,
\label{eq_xfmr_ex2_4}
\end{equation}
which is obtained by differentiating
Eq.~\ref{eq_xfmr_ex2_3}. Although this is a very simple idea, assembling
Eq.~\ref{eq_xfmr_ex2_4} from the netlist information requires a systematic approach which will
work generally. The following steps constitute one such approach.
\begin{list}{\arabic{cntr1}.}{\usecounter{cntr1}}
 \item
  Find inductors in series with each of the two coils ($p$ and $s$) using an algorithm similar
  to the shorting algorithm of Sec.~\ref{sec_ind_series}; the only difference here is that we
  are looking for a series connection between a transformer coil and an inductor rather than
  between two inductors. We find $L_1$ to be in series with the $p$ coil and $L_2$ to be in
  series with the $s$ coil.
 \item
  Write the transformer equation relating $i_{pp}$ and $i_{sp}$, viz.,
  \begin{equation}
  N_p i_{pp} + N_s i_{sp} = 0.
  \label{eq_xfmr_ex2_5}
  \end{equation}
 \item
  For each term in the above equation, replace the coil current with $i_L$ of the inductor
  (if any) in series with the coil, taking into account the relative directions of the coil
  current and the inductor current. For the circuit of Fig.~\ref{fig_xfmr_2},
  $i_{pp}$ gets replaced with $i_{L1}$ and
  $i_{sp}$ with $-i_{L2}$, leading to
  \begin{equation}
  N_p i_{L1} - N_s i_{L2} = 0.
  \label{eq_xfmr_ex2_6}
  \end{equation}
 \item
  If the resulting equation has only $i_L$ terms~-- which is true about Eq.~\ref{eq_xfmr_ex2_6}~--
  we replace each $i_L$ with the corresponding $i_{Ld}$ and obtain the desired equation
  (Eq.~\ref{eq_xfmr_ex2_4}).
\end{list}

\subsubsection{Transformer circuits: example 3}
Consider the circuit shown in Fig.~\ref{fig_xfmr_3} in which inductors $L_2$, $L_3$ are
directly in series, and their currents are also related to the current through $L_1$ on the
other side of the transformer. Of the three inductor currents, only one of them can be
independently assigned. A systematic procedure for writing the inductor CTD equations is
given below.
\label{sec_xfmr_ex_3}
\begin{figure}[!ht]
\centering
\scalebox{0.9}{\includegraphics{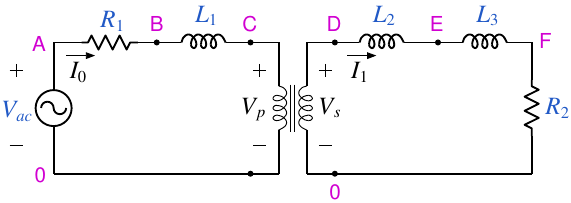}}
\vspace*{-0.2cm}
\caption{Transformer circuit example 3. The $p$ nodes of
$L_1$, $L_2$, $L_3$ are B, D, E, respectively.}
\label{fig_xfmr_3}
\end{figure}
\begin{list}{\arabic{cntr1}.}{\usecounter{cntr1}}
 \item
  Using the shorting algorithm, we find that $L_2$ and $L_3$ are directly in series.
  We now have two sets of inductors: $(L_1)$ and $(L_2,\,L_3)$. In the second set,
  we select one of the inductors as the ``first" (F) inductor, and the remaining~--
  in this case only one~-- as the ``non-first" (NF) inductor. For example, $L_2$ could
  be type F, and $L_3$ would then be type NF. In the first set, there is only one inductor
  ($L_1$), and it gets type F.
 \item
  We now look for series connections between the F inductors and the transformer coils and
  find that $L_1$ is in series with the $p$ coil, and $L_2$ is in series with the $s$ coil.
 \item
  Following the procedure described for the previous example, we find that the currents of
  the F inductors $L_1$ and $L_2$ are related by the transformer current equation. We get the
  following relationship.
  \begin{equation}
  N_p i_{L1d} - N_s i_{L2d} = 0.
  \label{eq_xfmr_ex3_1}
  \end{equation}
  We categorise one of the two F inductors ($L_1$ and $L_2$) as ``first-first" (FF)
  and the other inductor as ``first-non-first" (FNF). For example, we may select $L_1$ to be
  FF, which makes $L_2$ FNF.
 \item
  After the above step, if there are any other F inductors, we mark them as FF. In the present
  example, there are no inductors in this category.
 \item
  We now have three types of inductors: FF ($L_1$), FNF ($L_2$), and NF ($L_3$). For the FF
  inductors, we assign $i_L$ (i.e., the corresponding branch current) independently:
  \begin{equation}
  I_0 = i_{L1}^{(n+1)}.
  \label{eq_xfmr_ex3_2}
  \end{equation}
 \item
  For each NF inductor, we relate its $i_{Ld}$ to that of the corresponding FF inductor. For
  the inductor $L_3$, we get
  \begin{equation}
  i_{L3d} = i_{L2d}.
  \label{eq_xfmr_ex3_3}
  \end{equation}
\end{list}
Eqs.~\ref{eq_xfmr_ex3_1}-\ref{eq_xfmr_ex3_3} are the desired inductor equations.

\subsubsection{Transformer circuits: example 4}
\label{sec_xfmr_ex_4}
Fig.~\ref{fig_xfmr_4} shows a circuit with two coupled transformers.
\begin{figure}[!ht]
\centering
\scalebox{0.9}{\includegraphics{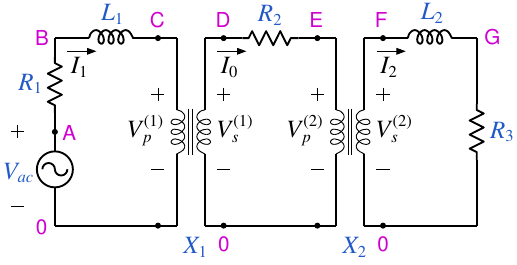}}
\vspace*{-0.2cm}
\caption{Transformer circuit example 4. The $p$ nodes of
$L_1$, $L_2$ are B, F, respectively. $X_1$ and $X_2$ denote the two
transformers.}
\label{fig_xfmr_4}
\end{figure}
There are two inductors $L_1$ and $L_2$ which are not directly connected in series,
but their currents are linearly related:
\begin{equation}
i_{L2} = \displaystyle\frac{N_p^{(1)}}{N_s^{(1)}} \times \displaystyle\frac{N_p^{(2)}}{N_s^{(2)}} \times i_{L1},
\label{eq_xfmr_ex4_1}
\end{equation}
where the superscripts $(1)$ and $(2)$ correspond to transformers $X_1$, $X_2$, respectively.
Since $i_{L1}$ and $i_{L2}$ are not independent, we should write the following equations:
\begin{equation}
I_1 = i_{L1}^{(n+1)}~{\textrm{or}}~I_2 = i_{L2}^{(n+1)},
\label{eq_xfmr_ex4_2}
\end{equation}
\begin{equation}
i_{L2d} = \displaystyle\frac{N_p^{(1)}}{N_s^{(1)}} \times \displaystyle\frac{N_p^{(2)}}{N_s^{(2)}} \times i_{L1d}.
\label{eq_xfmr_ex4_3}
\end{equation}
A systematic procedure to arrive at Eq.~\ref{eq_xfmr_ex4_3} is the following.
\begin{list}{\arabic{cntr1}.}{\usecounter{cntr1}}
 \item
  Write the current equations for the two transformers, $X_1$ and $X_2$.
  \begin{equation}
  N_p^{(1)} i_{pp}^{(1)} + N_s^{(1)} i_{sp}^{(1)} = 0,
  \label{eq_xfmr_ex4_4}
  \end{equation}
  \begin{equation}
  N_p^{(2)} i_{pp}^{(2)} + N_s^{(2)} i_{sp}^{(2)} = 0.
  \label{eq_xfmr_ex4_5}
  \end{equation}
 \item
  Use the shorting algorithm to check series connections between the transformer coils and inductors.
  We find that the $p$ coil of $X_1$ is in series with $L_1$, and the $s$ coil of $X_2$ is in series
  with $L_2$.
 \item
  Replace $i_{pp}^{(1)}$ with $i_{L1}$ in Eq.~\ref{eq_xfmr_ex4_4} and
  $i_{sp}^{(2)}$ with $(-i_{L2})$ in Eq.~\ref{eq_xfmr_ex4_5}. We now have
  \begin{equation}
  N_p^{(1)} i_{L1} + N_s^{(1)} i_{sp}^{(1)} = 0,
  \label{eq_xfmr_ex4_6}
  \end{equation}
  \begin{equation}
  N_p^{(2)} i_{pp}^{(2)} - N_s^{(2)} i_{L2} = 0.
  \label{eq_xfmr_ex4_7}
  \end{equation}
 \item
  Use the shorting algorithm to check series connections between transformer coils. We find that
  the $s$ coil of $X_1$ is in series with the $p$ coil of $X_2$ which implies
  \begin{equation}
  i_{sp}^{(1)} = -i_{pp}^{(2)}.
  \label{eq_xfmr_ex4_8}
  \end{equation}
 \item
  Use the above relationship to eliminate
  $i_{sp}^{(1)}$, $i_{pp}^{(2)}$ from
  Eqs.~\ref{eq_xfmr_ex4_6} and \ref{eq_xfmr_ex4_7} to get
  \begin{equation}
  N_p^{(1)} N_p^{(2)} i_{L1} - N_s^{(1)} N_s^{(2)} i_{L2} = 0.
  \label{eq_xfmr_ex4_9}
  \end{equation}
  Differentiating Eq.~\ref{eq_xfmr_ex4_9} with respect to time gives us the desired
  equation (Eq.~\ref{eq_xfmr_ex4_1}).
\end{list}

\subsubsection{Transformer circuits: example 5}
\label{sec_xfmr_ex_5}
We now consider a circuit with 1:2 transformers, as shown in Fig.~\ref{fig_xfmr_5}.
\begin{figure}[!ht]
\centering
%\hspace*{-0.5cm}\scalebox{0.9}{\includegraphics{xfmr_5.pdf}}
\scalebox{0.9}{\includegraphics{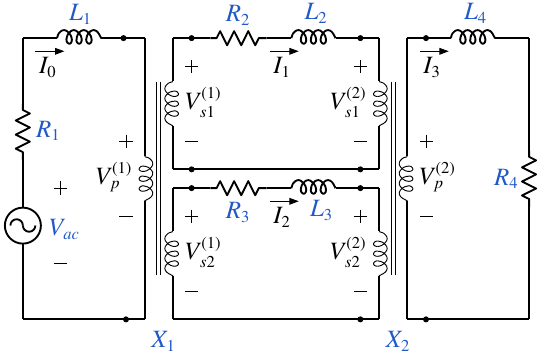}}
\vspace*{-0.2cm}
\caption{Transformer circuit example 5. The inductor currents
$i_{L1}$, $i_{L2}$, $i_{L3}$, $i_{L4}$ are in the same direction as
the branch currents, $I_0$, $I_1$, $I_2$, $I_3$, respectively.}
\label{fig_xfmr_5}
\end{figure}
For convenience, we will consider the side with one coil to be the primary side
and the other side with two coils to be the secondary side. In this circuit,
we have four inductors whose currents are not independent but are linked through
the transformer current equations. The following procedure yields the desired set
of inductor CTD equations.
\begin{list}{\arabic{cntr1}.}{\usecounter{cntr1}}
 \item
  Start with the transformer current equations:
  \begin{equation}
  N_p^{(1)} i_{pp}^{(1)} + N_{s1}^{(1)} i_{s1p}^{(1)} + N_{s2}^{(1)} i_{s2p}^{(1)} = 0,
  \label{eq_xfmr_ex5_1}
  \end{equation}
  \begin{equation}
  N_p^{(2)} i_{pp}^{(2)} + N_{s1}^{(2)} i_{s1p}^{(2)} + N_{s2}^{(2)} i_{s2p}^{(2)} = 0.
  \label{eq_xfmr_ex5_2}
  \end{equation}
 \item
  Replace the coil current with $i_L$ of the F inductor in series with that coil. This gives
  \begin{equation}
  N_p^{(1)} i_{L1} - N_{s1}^{(1)} i_{L2} - N_{s2}^{(1)} i_{L3} = 0,
  \label{eq_xfmr_ex5_3}
  \end{equation}
  \begin{equation}
  -N_p^{(2)} i_{L4} + N_{s1}^{(2)} i_{L2} + N_{s2}^{(2)} i_{L3} = 0.
  \label{eq_xfmr_ex5_4}
  \end{equation}
 \item
  Since both the above equations have only $i_L$ terms,
  we replace each $i_L$ with the corresponding $i_{Ld}$ and add the resulting equations
  to the set of circuit equations to be solved:
  \begin{equation}
  N_p^{(1)} i_{L1d} - N_{s1}^{(1)} i_{L2d} - N_{s2}^{(1)} i_{L3d} = 0,
  \label{eq_xfmr_ex5_5}
  \end{equation}
  \begin{equation}
  -N_p^{(2)} i_{L4d} + N_{s1}^{(2)} i_{L2d} + N_{s2}^{(2)} i_{L3d} = 0.
  \label{eq_xfmr_ex5_6}
  \end{equation}
 \item
  We have four inductors and only two inductor CTD equations so far. We need to write
  two branch current assignment equations. One of the choices we can make is
  \begin{equation}
  I_0 = i_{L1},
  \label{eq_xfmr_ex5_7}
  \end{equation}
  \begin{equation}
  I_1 = i_{L2}.
  \label{eq_xfmr_ex5_8}
  \end{equation}
\end{list}

\subsubsection{Transformer circuits: example 6}
\label{sec_xfmr_ex_6}
If $L_2$ in our previous example (Fig.~\ref{fig_xfmr_5}) is removed (shorted),
we get the circuit shown in Fig.~\ref{fig_xfmr_6}.
\begin{figure}[!ht]
\centering
\scalebox{0.9}{\includegraphics{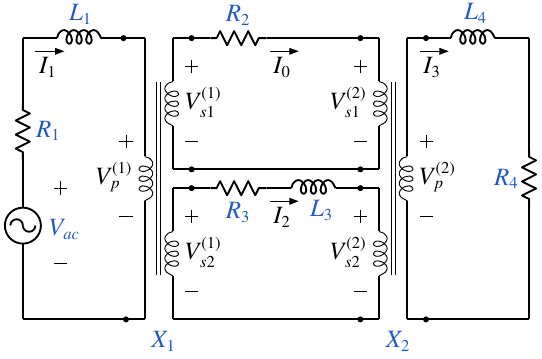}}
\vspace*{-0.2cm}
\caption{Transformer circuit example 6. The inductor currents
$i_{L1}$, $i_{L3}$, $i_{L4}$ are in the same direction as
the branch currents, $I_1$, $I_2$, $I_3$, respectively.}
\label{fig_xfmr_6}
\end{figure}
We now have three inductors, $L_1$, $L_3$, $L_4$, and it is not obvious if their
currents can be independently assigned. The following procedure can be used to arrive
at the desired equations.
\begin{list}{\arabic{cntr1}.}{\usecounter{cntr1}}
 \item
  Start with the transformer current equations
  (Eqs.~\ref{eq_xfmr_ex5_1}, \ref{eq_xfmr_ex5_2}) and for each coil replace the coil current
  with $i_L$ of the F inductor~-- if any~-- in series with that coil. This gives
  \begin{equation}
  N_p^{(1)} i_{L1} + N_{s1}^{(1)} i_{s1p}^{(1)} - N_{s2}^{(1)} i_{L3} = 0,
  \label{eq_xfmr_ex6_1}
  \end{equation}
  \begin{equation}
  -N_p^{(2)} i_{L4} + N_{s1}^{(2)} i_{s1p}^{(2)} + N_{s2}^{(2)} i_{L3} = 0.
  \label{eq_xfmr_ex6_2}
  \end{equation}
 \item
  Use $i_{s1p}^{(1)} \,$=$\, -i_{s1p}^{(2)}$ to combine
  Eqs.~\ref{eq_xfmr_ex6_1} and \ref{eq_xfmr_ex6_2}:
  \begin{multline}
  N_p^{(1)} N_{s1}^{(2)} i_{L1}
  + \left(N_{s1}^{(1)} N_{s2}^{(2)} - N_{s2}^{(1)} N_{s1}^{(2)}\right)\,i_{L3}
  \\
  - N_{s1}^{(1)} N_{p}^{(2)} i_{L4}
  = 0.
  \label{eq_xfmr_ex6_3}
  \end{multline}
 \item
  Since Eq.~\ref{eq_xfmr_ex6_3} has only $i_L$ terms,
  we replace each $i_L$ with the corresponding $i_{Ld}$ and add the resulting equation
  to the set of circuit equations to be solved:
  \begin{multline}
  N_p^{(1)} N_{s1}^{(2)} i_{L1d}
  + \left(N_{s1}^{(1)} N_{s2}^{(2)} - N_{s2}^{(1)} N_{s1}^{(2)}\right)\,i_{L3d}
  \\
  - N_{s1}^{(1)} N_{p}^{(2)} i_{L4d}
  = 0.
  \label{eq_xfmr_ex6_4}
  \end{multline}
 \item
  Since we have three inductors and one inductor CTD equation, we write
  two branch current assignment equations. For example,
  \begin{equation}
  I_1 = i_{L1},
  \label{eq_xfmr_ex6_5}
  \end{equation}
  \begin{equation}
  I_2 = i_{L3}.
  \label{eq_xfmr_ex6_6}
  \end{equation}
\end{list}

\subsubsection{Transformer circuits: example 7}
\label{sec_xfmr_ex_7}
Consider the circuit shown in Fig.~\ref{fig_xfmr_7} which has two inductors, $L_1$ and $L_4$.
\begin{figure}[!ht]
\centering
\scalebox{0.9}{\includegraphics{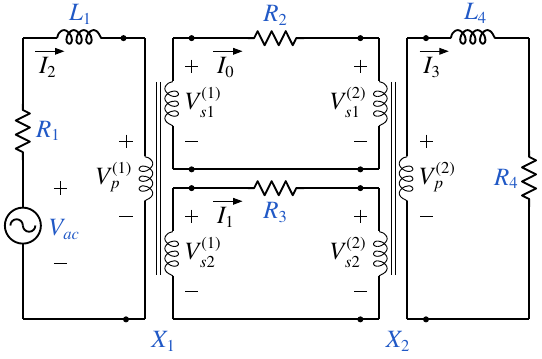}}
\vspace*{-0.2cm}
\caption{Transformer circuit example 7. The inductor currents
$i_{L1}$, $i_{L4}$ are in the same direction as
the branch currents, $I_2$, $I_3$, respectively.}
\label{fig_xfmr_7}
\end{figure}
At first glance, they appear to be coupled through the transformer current equations.
However, in order to firmly conclude if $i_{L1}$ and $i_{L4}$ can be independently
assigned, a systematic approach is called for. To this end, we perform the following
steps.
\begin{list}{\arabic{cntr1}.}{\usecounter{cntr1}}
 \item
  Start with the transformer current equations and replace coil currents with
  inductor currents as discussed in the previous examples. This gives
  \begin{equation}
  N_p^{(1)} i_{L1} + N_{s1}^{(1)} i_{s1p}^{(1)} + N_{s2}^{(1)} i_{s2p}^{(1)} = 0,
  \label{eq_xfmr_ex7_1}
  \end{equation}
  \begin{equation}
  -N_p^{(2)} i_{L4} + N_{s1}^{(2)} i_{s1p}^{(2)} + N_{s2}^{(2)} i_{s2p}^{(2)} = 0.
  \label{eq_xfmr_ex7_2}
  \end{equation}
 \item
  Use $i_{s1p}^{(1)} \,$=$\, -i_{s1p}^{(2)}$ to combine
  Eqs.~\ref{eq_xfmr_ex7_1} and \ref{eq_xfmr_ex7_2}:
  \begin{multline}
   N_{p}^{(1)} N_{s1}^{(2)} i_{L1}
  +N_{s2}^{(1)} N_{s1}^{(2)} i_{s2p}^{(1)}
  -N_{s1}^{(1)} N_{p}^{(2)} i_{L4}
  \\
  +N_{s1}^{(1)} N_{s2}^{(2)} i_{s2p}^{(2)}
   = 0.
  \label{eq_xfmr_ex7_3}
  \end{multline}
 \item
  Use $i_{s2p}^{(1)} \,$=$\, -i_{s2p}^{(2)}$ to simplify
  Eq.~\ref{eq_xfmr_ex7_3} as
  \begin{multline}
   N_{p}^{(1)} N_{s1}^{(2)} i_{L1}
  -N_{s1}^{(1)} N_{p}^{(2)} i_{L4}
   \\
  +\left(
   N_{s2}^{(1)} N_{s1}^{(2)}
  -N_{s1}^{(1)} N_{s2}^{(2)}
   \right)
   i_{s2p}^{(1)}
   = 0.
  \label{eq_xfmr_ex7_4}
  \end{multline}
 \item
  If $ N_{s2}^{(1)} N_{s1}^{(2)} \,\neq\, N_{s1}^{(1)} N_{s2}^{(2)} $, we have a ``mixed"
  equation containing inductor currents and transformer coil currents. We conclude that
  $i_{L1}$ and $i_{L4}$ can be independently equated to the respective branch currents, i.e.,
  \begin{equation}
  I_2 = i_{L1},
  \label{eq_xfmr_ex7_5}
  \end{equation}
  \begin{equation}
  I_3 = i_{L4}.
  \label{eq_xfmr_ex7_6}
  \end{equation}
  If $ N_{s2}^{(1)} N_{s1}^{(2)} \,$=$\, N_{s1}^{(1)} N_{s2}^{(2)} $,
  Eq.~\ref{eq_xfmr_ex7_4} simplifies to
  \begin{equation}
   N_{p}^{(1)} N_{s1}^{(2)} i_{L1}
  -N_{s1}^{(1)} N_{p}^{(2)} i_{L4}
   = 0.
  \label{eq_xfmr_ex7_7}
  \end{equation}
  In this case, $i_{L1}$ and $i_{L4}$ are not independent.
  Differentiating Eq.~\ref{eq_xfmr_ex7_7} with respect to time, we obtain one of 
  the desired equations:
  \begin{equation}
   N_{p}^{(1)} N_{s1}^{(2)} i_{L1d}
  - N_{s1}^{(1)} N_{p}^{(2)} i_{L4d}
   = 0,
  \label{eq_xfmr_ex7_8}
  \end{equation}
  and we should write the branch current assignment equation for only one of the
  two inductors, e.g., Eq.~\ref{eq_xfmr_ex7_5}.
\end{list}

\section{Conclusions}
\label{sec_conclusions}
In this paper, we have pointed out some issues arising in using explicit integration
schemes for simulation of power electronic circuits with ideal switch models. In
particular, the following topics have been addressed:
(a)\,treatment of circuits involving switch loops and isolated sections,
(b)\,checking series connection of inductors,
(c)\,conduction path for inductors,
(d)\,inductor equations in the presence of transformers.
Systematic algorithms have been presented to address the various challenges posed
in assembling the circuit equations. Application of the new methods and algorithms
is illustrated with the help of suitable examples.

The techniques presented in this paper have been incorporated in the open-source
circuit simulator GSEIM\,\cite{gseimgithub}. In parallel, several improvements in the
graphics interface of GSEIM are also being undertaken. The new version of GSEIM, when
made available in the public domain, is expected to provide a versatile, open-source alternative
for simulation of power electronic circuits.

\bibliographystyle{IEEEtran}
\bibliography{ref4a}

\begin{thebibliography}{1}
\providecommand{\url}[1]{#1}
\csname url@samestyle\endcsname
\providecommand{\newblock}{\relax}
\providecommand{\bibinfo}[2]{#2}
\providecommand{\BIBentrySTDinterwordspacing}{\spaceskip=0pt\relax}
\providecommand{\BIBentryALTinterwordstretchfactor}{4}
\providecommand{\BIBentryALTinterwordspacing}{\spaceskip=\fontdimen2\font plus
\BIBentryALTinterwordstretchfactor\fontdimen3\font minus
  \fontdimen4\font\relax}
\providecommand{\BIBforeignlanguage}[2]{{%
\expandafter\ifx\csname l@#1\endcsname\relax
\typeout{** WARNING: IEEEtran.bst: No hyphenation pattern has been}%
\typeout{** loaded for the language `#1'. Using the pattern for}%
\typeout{** the default language instead.}%
\else
\language=\csname l@#1\endcsname
\fi
#2}}
\providecommand{\BIBdecl}{\relax}
\BIBdecl

\bibitem{plecs}
\BIBentryALTinterwordspacing
{PLECS 4.6}. [Online]. Available: \url{https://www.plexim.com/products/plecs}
\BIBentrySTDinterwordspacing

\bibitem{elex1}
{\mbox{M.B. Patil}} and {\mbox{V.V.S. Pavan Kumar Hari}}, ``Circuit simulation
  using explicit methods,'' \emph{arXiv preprint arXiv:2301.04595}, 2023.

\bibitem{elex2}
{\mbox{M.B. Patil}}, ``Circuit simulation using explicit methods: singular
  matrix issues,'' \emph{arXiv preprint arXiv:2306.13489}, 2023.

\bibitem{ngspice}
P.~Nenzi, ``{NGSPICE} circuit simulator release 26, 2014.''

\bibitem{gseimgithub}
\BIBentryALTinterwordspacing
{GSEIM}. [Online]. Available: \url{https://github.com/gseim/gseim}
\BIBentrySTDinterwordspacing

\end{thebibliography}

\end{document}